# A Framework for Feature Discovery in Intracranial
## Pressure Monitoring Data Using Neural Network Attention

Jonathan D. Socha, Seyed F. Maroufi, Dipankar Biswas, Richard Um, Aruna S. Rao, Mark G. Luciano

*Abstract*—We present a novel framework for analyzing intracranial pressure monitoring data by applying interpretability principles. Intracranial pressure monitoring data was collected from 60 patients at Johns Hopkins. The data was segmented into individual cardiac cycles. A convolutional neural network was trained to classify each cardiac cycle into one of seven body positions. Neural network attention was extracted and was used to identify regions of interest in the waveform. Further directions for exploration are identified. This framework provides an extensible method to further understand the physiological and clinical underpinnings of the intracranial pressure waveform, which could lead to better diagnostic capabilities for intracranial pressure monitoring.

*Corresponding author: Jonathan D. Socha.*

Jonathan D. Socha, Seyed F. Maroufi, Dipankar Biswas, Richard Um, and Mark G. Luciano are with the Department of Neurosurgery, Johns Hopkins University School of Medicine, Baltimore, MD 21205 USA.

Aruna Rao is with the Department of Neurology, Johns Hopkins University School of Medicine, Baltimore, MD 21205 USA.

## I. INTRODUCTION

THE intracranial pressure (ICP) waveform (Fig. 1) results from the complex interaction of various physiological systems involved in blood flow and blood volume, cerebrospinal fluid production, absorption and movement, brain elasticity, and respiration [1]. Although the ICP waveform has been characterized, the physiological generators of the waveform are still not fully understood at a mechanistic level [2]. The ICP is implicated as an indicator and result of various disorders including cerebrospinal fluid disorders like hydrocephalus, Chiari malformation, and traumatic brain injuries.

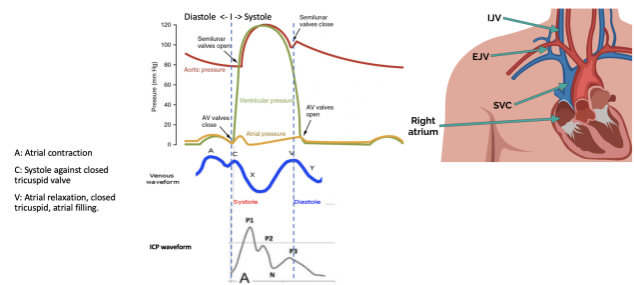

Jugular venous pressure (JVP) curve

A: Atrial contraction
C: Systole against closed tricuspid valve
V: Atrial relaxation, closed tricuspid, atrial filling

**Fig. 1.** Known physiological underpinnings of the ICP at the level of the cardiac cycle.

Intracranial pressure monitoring (ICPm) is conducted as part of the diagnostic pipeline for cerebrospinal fluid disorders. When patients present with nonspecific complaints like headaches, gait changes, and urinary incontinence, a battery of tests is performed. If these tests do not lead to a conclusive diagnosis but do indicate a suspected diagnosis, the patient undergoes ICPm. Intracranial pressure probes are inserted through a burr hole drilled in the patient's skull. These probes measure the pressure in the brain directly from the intracranial space or from the brain using a pressure transducer. Patients may be monitored for multiple days and may undergo repeat monitoring sessions over time. Other data points may also be captured concurrently with ICP, such as heart rate, EKG, PPG, respirations, and $SaO_2$. Clinical personnel can label the data as it is acquired or post-acquisition, noting environmental or behavioral factors. For example: 'patient coughed,' 'patient got up to use the restroom,' 'monitor disconnected,' 'patient sleeping.'

In addition, more targeted measurements can be made via specific testing regimens. One of these regimens is positional testing. Patients are monitored at different body positions, where the patient's head is positioned relative to his or her torso at 0° (supine), 10°, 20°, 30°, 45°, 60°, and 90° (sitting up). The patient may spend 3 – 5 minutes at each position, depending on physician preference. As the patient moves through positions, his or her ICP and brain compliance are expected to change, with greater elevation associated with increased compliance. Brain compliance is the inverse of elastance and refers to the ability of the brain to compensate for changes in volume without an associated increase in pressure. A patient with low compliance will see a greater increase in pressure for the same increase in volume than a patient with high compliance [3], [4], [5]. Once ICPm data is acquired, it is sometimes annotated and is then analyzed by a clinician(s) as a data point in the patient's clinical record to aid in a diagnosis. Extracted data points in the patient's chart may include, for each position, pressure ranges (mm Hg), average pressure, heart rate, and blood pressure. Diagnosis or outcome may be qualitative, such as 'positional testing demonstrates normal ICP,' and is generally taken as a single data point in a broader diagnostic or outcomes pathway.

The ICP contains information at multiple time scales, and these data are processed according to established practices known to clinicians to indicate physiological significance. At longer timescales, the average ICP is a dominant measure. Average ICP is a gross measurement that can indicate to a clinician whether the patient is hypotensive, hypertensive, or neither. At the level of individual cardiac cycles, the morphology of the waveform has features that have been discovered and are used to indicate information about the patient's brain compliance, venous system, and arterial system. One such feature is the P1/P2 ratio. P1 is commonly defined as the first inflection point in the cardiac cycle, and P2 is the second inflection point. A higher P2/P1 ratio has been noted to associate with lower compliance. Other examples of extracted features are Mean ICP Wave Amplitude and static ICP/mean ICP [6], Frequency/proportion of waves with a Mean Wave Amplitude greater than 5 mmHg [7], ICP Pulse Pressure Amplitude [6], Index of cerebrospinal compensatory reserve capacity [6], [8], ICP amplitude of the respiration induced waves [6], I/E ratio [9], Amplitude of slow vasogenic waves of the ICP signal [10], and Higher Harmonics Centroid [11].

However, the direct physiological and clinical correlates of waveform morphology are still unclear, i.e., what causes changes in waveform morphology under different physiological or clinical conditions. The usable features from the ICPm data are minimal and generally take the form of summary statistics. Much of the data cannot be used because the correlates of the waveform morphology and/or features are not well understood. As a result, the diagnostic utility of ICPm is limited. Because the ICP originates from diverse physiological processes, it is not a direct indicator like ECG or EEG that results from electrical signals. Feature discovery workflows have been studied to attempt to define clinically and physiologically relevant features of the ICP waveform at multiple timescales. The discovery workflows follow the general pattern of conducting a sweep of known features, identifying the most relevant features through a feature importance search, and further analyzing those features. Identifying novel features using this method is difficult. Univariate time series classification and feature extraction using classical machine learning is a mature field [12], [13]. In addition, controlled experiments would allow clinicians to more quickly identify and verify clinical and physiological waveform correlates, but it is difficult to justify targeted experimental interventions in patients. Another challenge is the curse of dimensionality: as the number of features of the data increases, the data becomes sparse. This affects classical machine learning pipelines. Deep learning overcomes the curse of dimensionality although the mechanism by which it does so is not well understood [14].

We developed a novel framework to identify the physiological and clinical factors underlying the ICP at the level of the cardiac cycle, using neural network attention, that aligns with previous work in mechanistic interpretability [15], [18]. Our dataset includes ICPm conducted for 377 patients. Of 373 patients, 18.2% were male and 81.8% female. Of 371 patients, age ranged from 15 to 78 years, with a mean of 40 and a standard deviation of 15. Of 357 patients, 44.5% of had an initial diagnosis of pseudotumor cerebri, 13.4% congenital hydrocephalus, 16.8% leak, 3.1% post-trauma/surgery, 4.8% chronic hydrocephalus, and 17.4% none. Of 331 patients, 29.6% had medication before monitoring, 50.2% had surgery, 24.5% had a blood patch, 64.4% prior lumbar puncture, 8.5% shunt tap. Of 344 patients, body mass index ranged from 14 to 66, with a mean of 31.8 and a standard deviation of 8.23. Of 372 patients, 40.3% had a suspected clinical diagnosis before ICPm of hypertensive, 41.4% hypotensive, 16.1% mixed, and 4.3% other. After ICPm, 12.4% had a clinical diagnosis of hypertensive, 34.4% hypotensive, 1.3% mixed, 52.2% normal, and 12.9% other. 250 of these patients underwent positional testing. We applied our analysis to 53 of these patients whose data was available. We used a separate cohort of 10 patients in an experimental cohort for data visualization. This 10-patient cohort was not included in the training, validation, or testing datasets.

## II. METHODS

Retrospective ICPm data was acquired from patients with hydrocephalus who underwent positional testing. During positional testing, patients were placed at different angles on a bed: 0° (supine), 10°, 20°, 30°, 45°, 60°, and 90° (upright). Data was acquired using a Codman ICP monitor with PowerLab at a sampling rate of 100 Hz. Data was exported to LabChart in .adicht format. ICP data was then extracted for each patient (Fig. 2). Patients were de-identified by assigning each monitoring session to a specific number; patients with multiple sessions were assigned multiple numbers. For each patient, positions were manually labeled by the clinician during or after testing. Labels were used to identify data segments for extraction. For each position, the entire positional testing period's data was extracted, except the beginning and end of the segment where the ICP changed from the previous position. Data was exported in .txt format and renamed position_patient (I00_01, I10_01, I00_02, etc.).

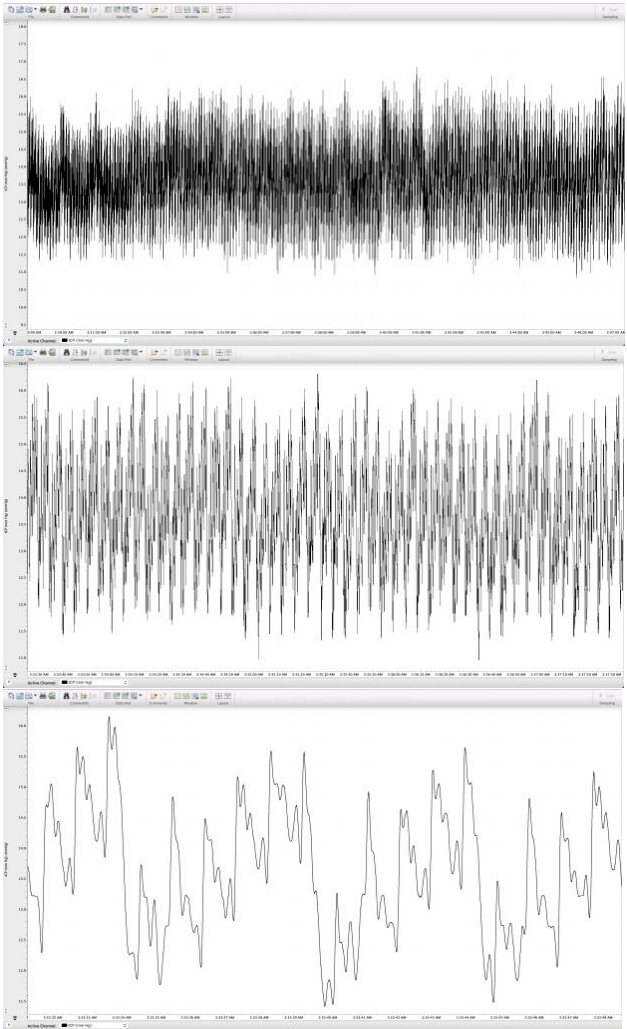

**Fig. 2.** Representative views of ICPm in LabChart at different time scales.

Each txt file was imported into Python and written to a dataframe using Pandas. Dataframes had two columns: time and pressure. NaN values were removed from both columns. Any dataframe with length < 100 was not processed. The data was normalized and detrended using statsmodels time series analysis tools (statsmodels.tsa.tstools detrend) with a detrending order of 10 (Fig. 3). Scipy's signal processing module (scipy.signal find_peaks) was used to identify peaks in the data (Fig. 4, Fig. 5). It was noted that peak detection was more effective at identifying troughs than peaks, so the algorithm was run on the negative of the detrended data. If length of peaks was < 2, the data was not processed. Distance was defined at 70 data points. The data was then segmented into individual cardiac cycles (Fig. 6). A new empty array was created. For each peak n to peak n + 1, if the length of the segment was > 20, the segment was interpolated to a target length of 100 using 1D linear interpolation and appended to this array. Each segment was labeled with its respective position by extracting the position label from the file name. The result of this process was an array of segments and an array of positions. These arrays were converted to

NumPy arrays: X (samples, channels, length) and y (position) with dimensions (182337, 1, 100) and (182337,). The number of segments per position was 12959 at 0°, 47895 at 10°, 23700 at 20°, 18877 at 30°, 25147 at 45°, 25886 at 60°, and 27873 at 90°.

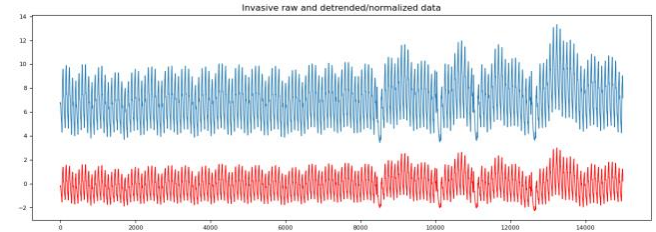

**Fig. 3.** Raw ICPm data (top) and normalized and detrended data (bottom).

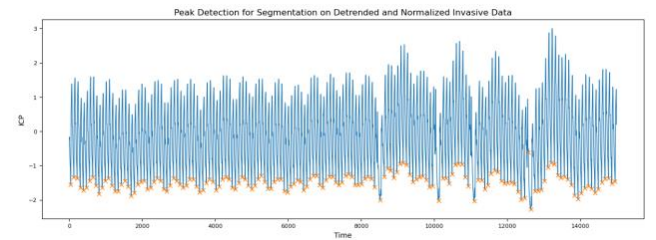

**Fig. 4.** Sample result of the peak-finding algorithm on detrended and normalized ICPm data. Each orange 'x' indicates a detected peak.

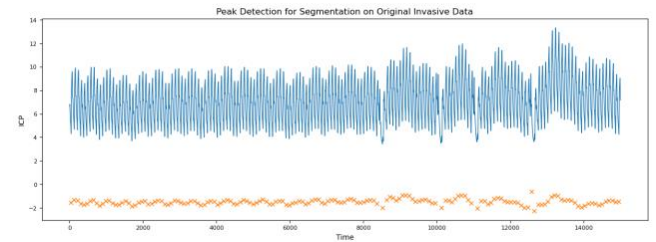

**Fig. 5.** Peak detection on the raw ICPm data.

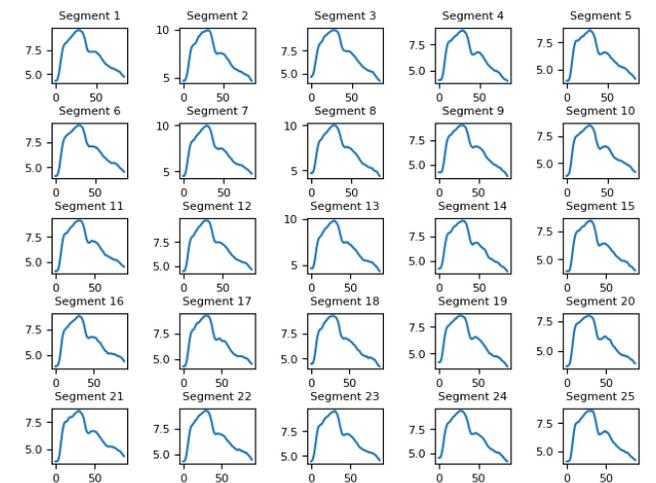

**Fig. 6.** 25 individual cardiac cycles segmented from ICPm data.

Distinct differences between positions are visually apparent when comparing the position averages across all patients (Fig. 7, Fig. 8). Regions of greatest difference appear to be from $0.2 - 0.4$ and $0.8 - 1.0$. The P2/P1 ratio increases with elevation, indicating lower compliance. However, this trend may not generalize to broader patient populations and may not be indicative of individual patient etiology. It may also be driven by an outlier patient or patients.

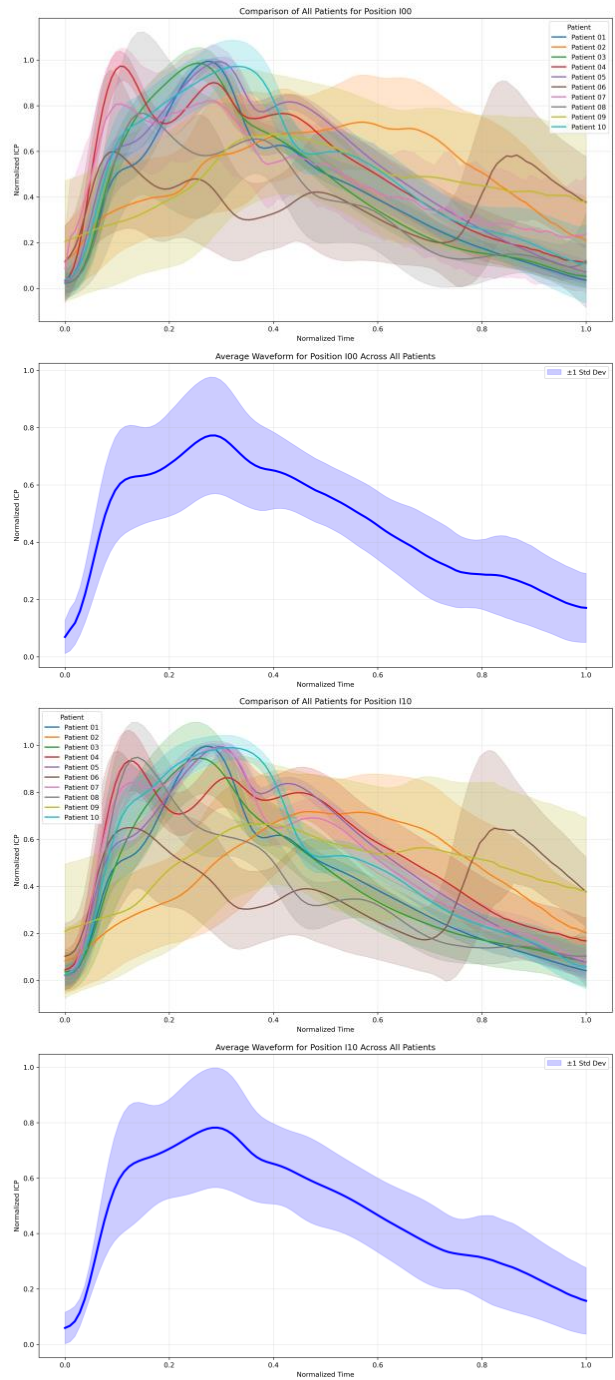

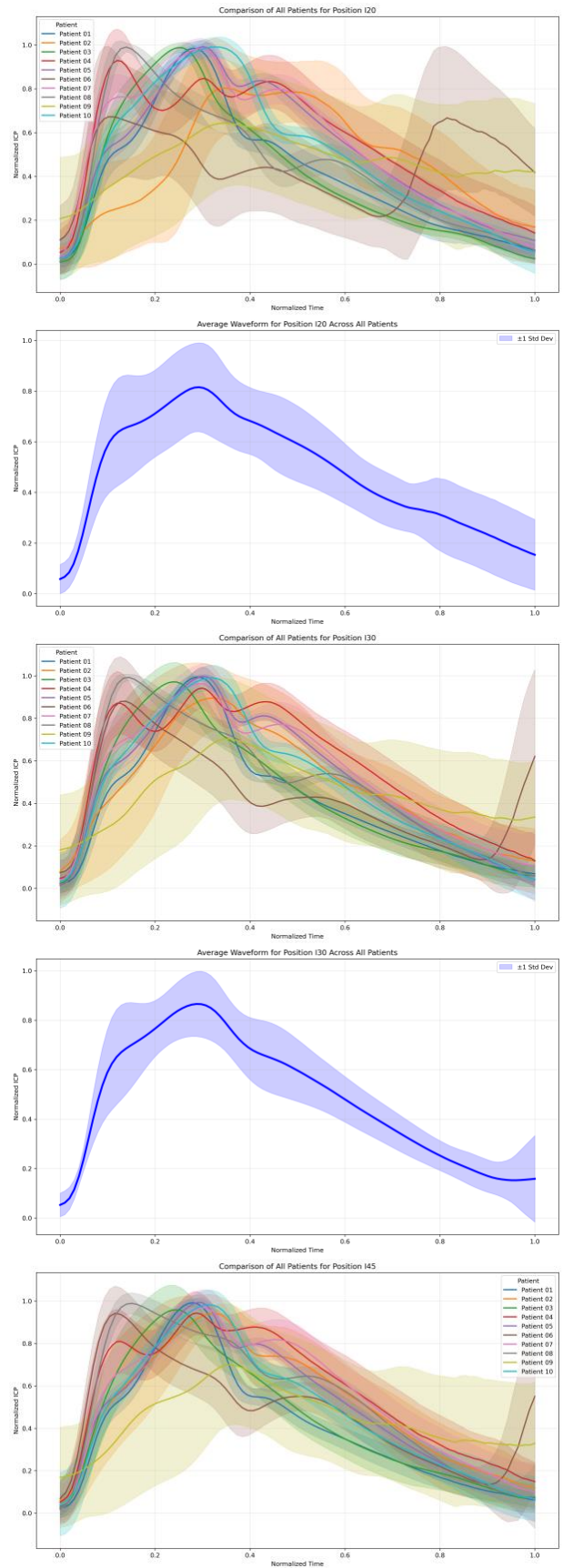

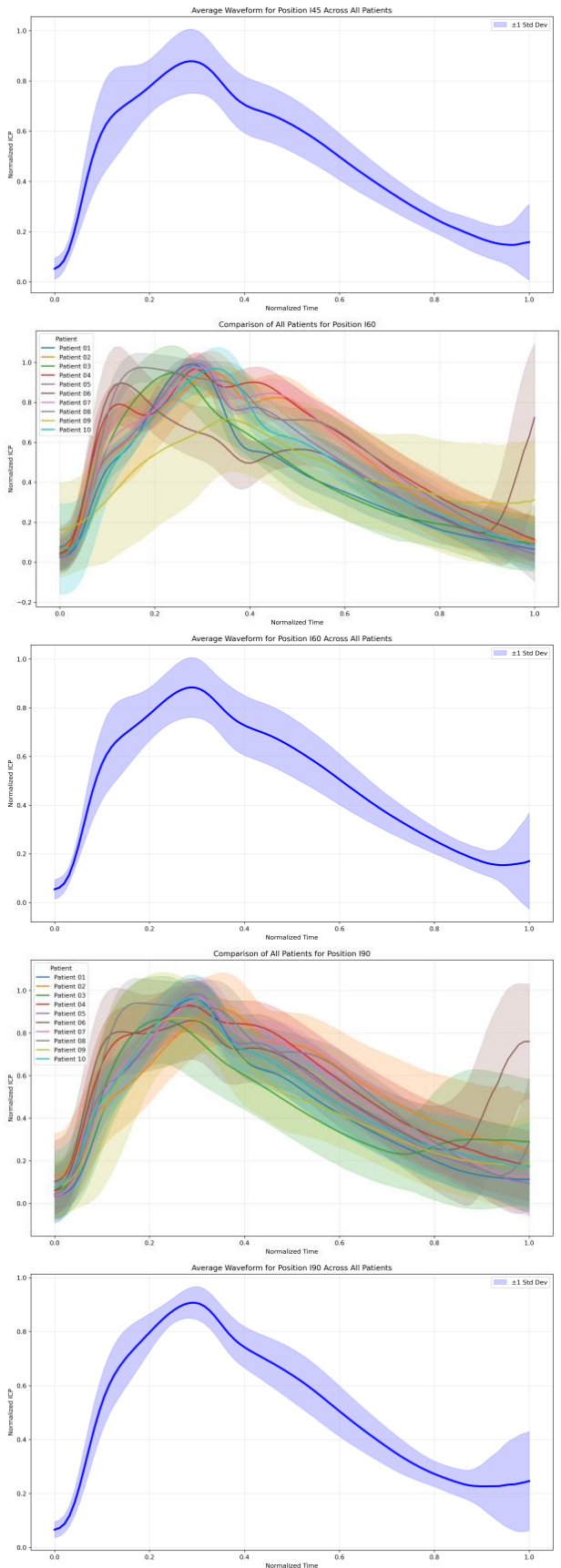

**Fig. 7.** Comparison of 10 patients over each position (top) and averages for each position (bottom).

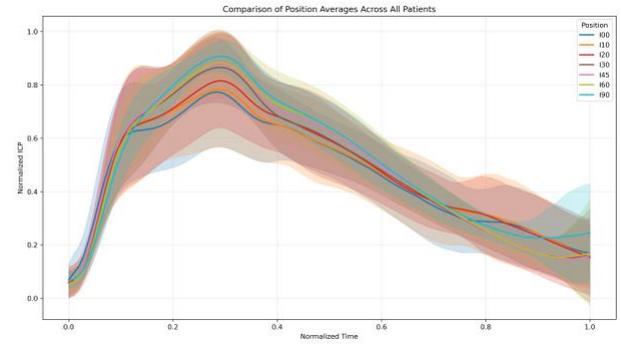

**Fig. 8.** Comparison of the averages of each position across 10 patients. Bold lines are averages, and shaded regions are +/- one standard deviation.

A comparison of all positions across each patient (Fig. 9) shows that intra-patient differences between positions vary. The shape of the cardiac cycle also varies significantly across patients.

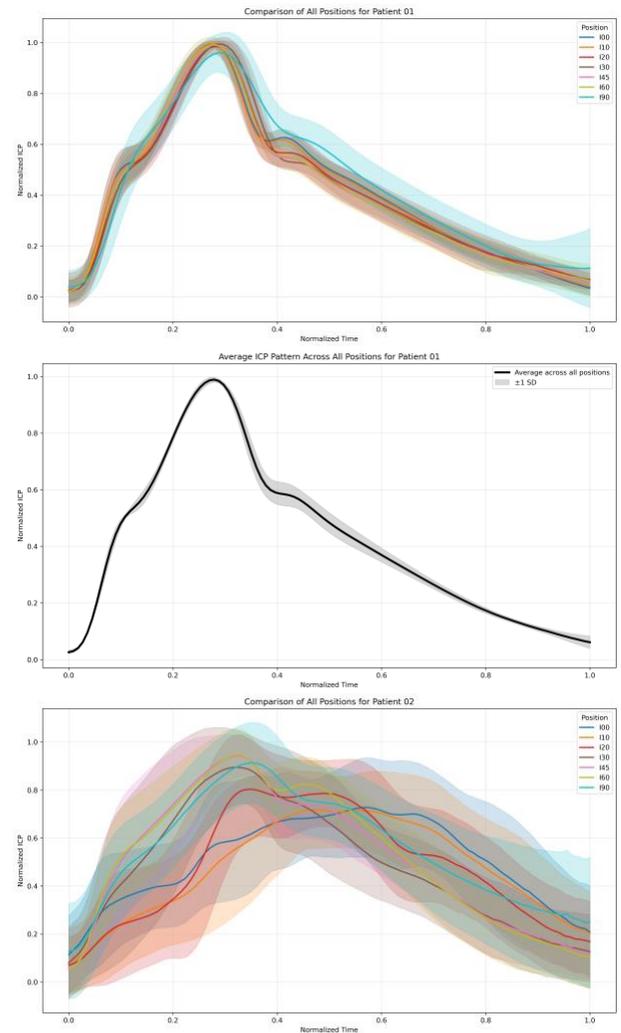

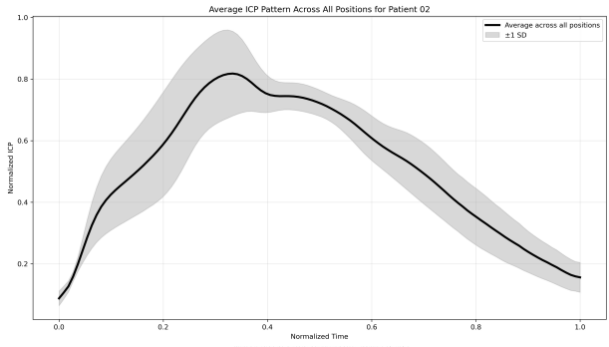

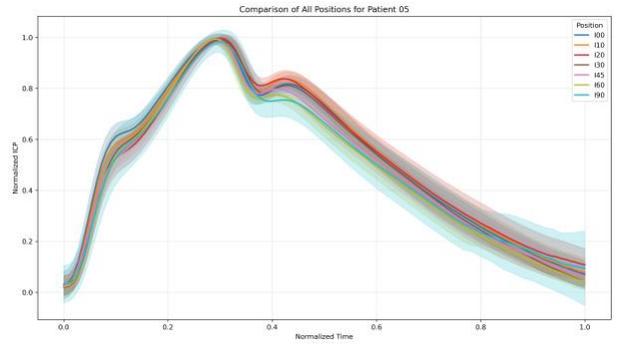

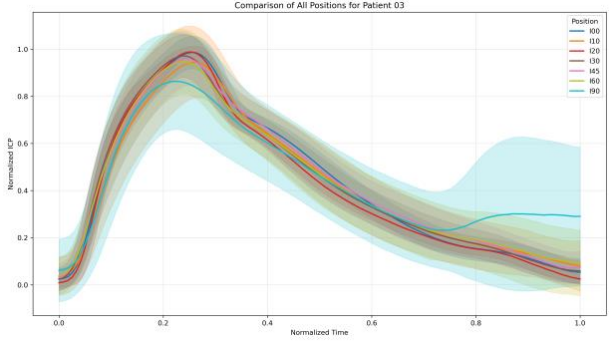

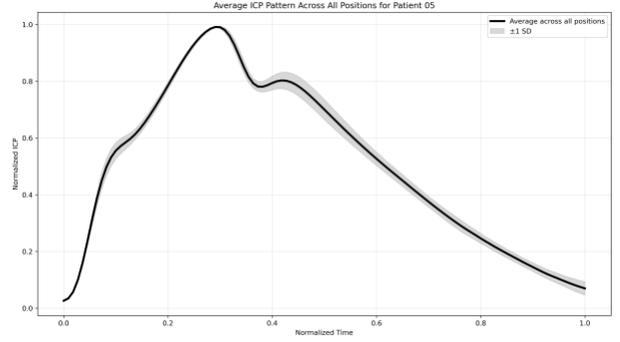

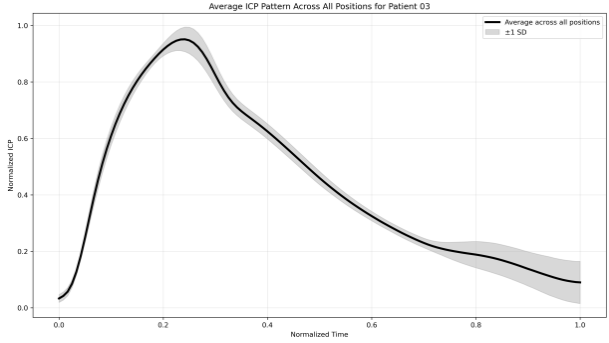

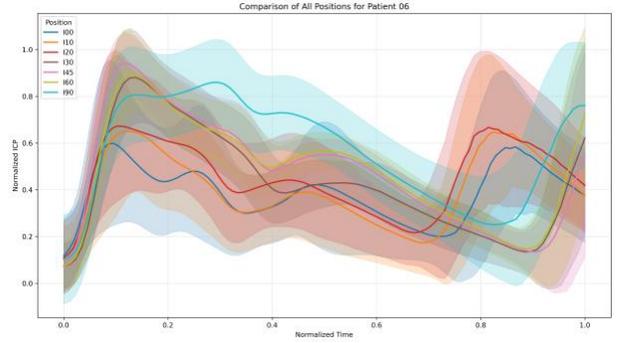

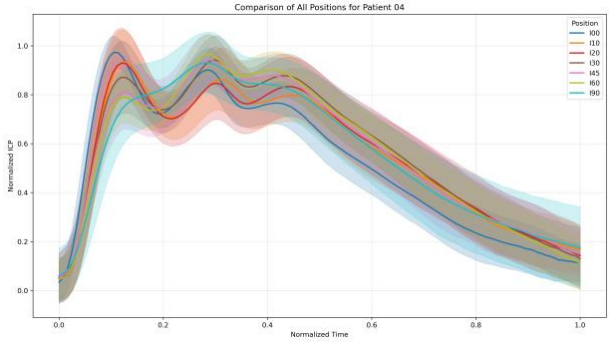

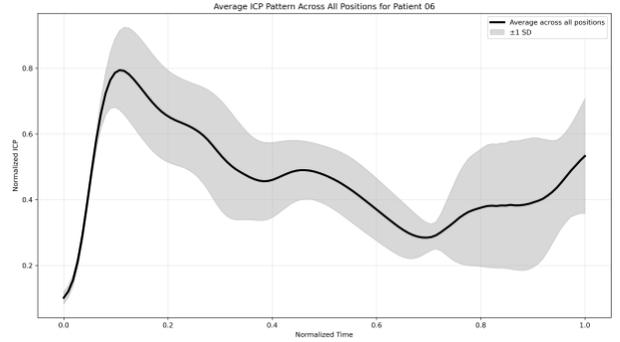

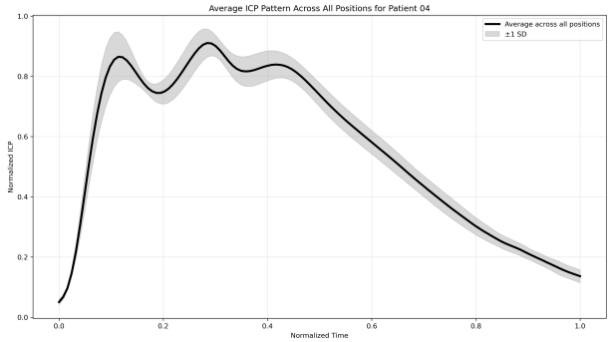

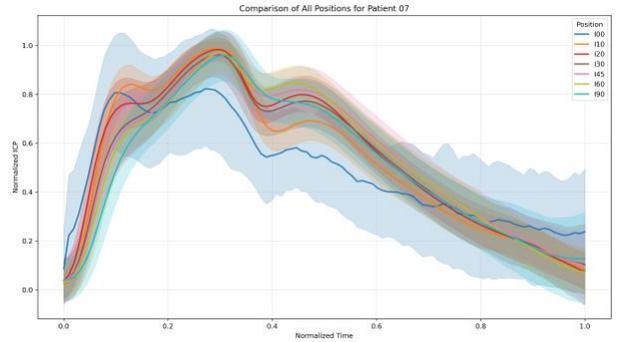

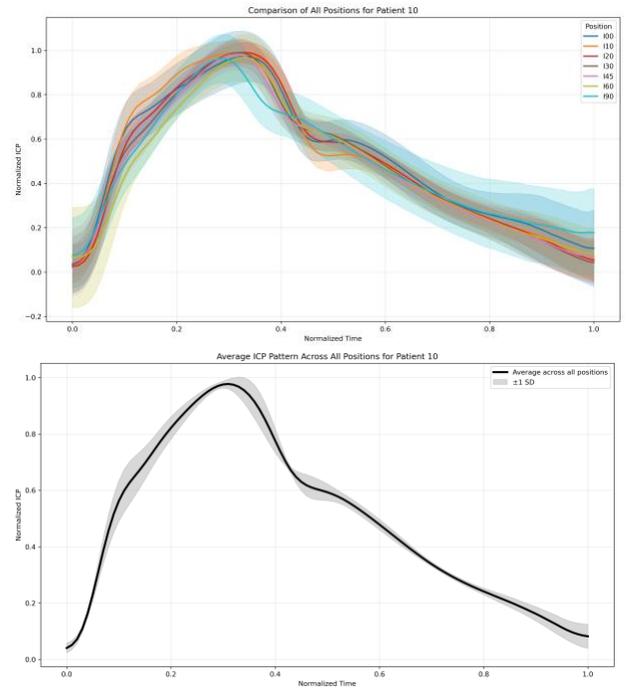

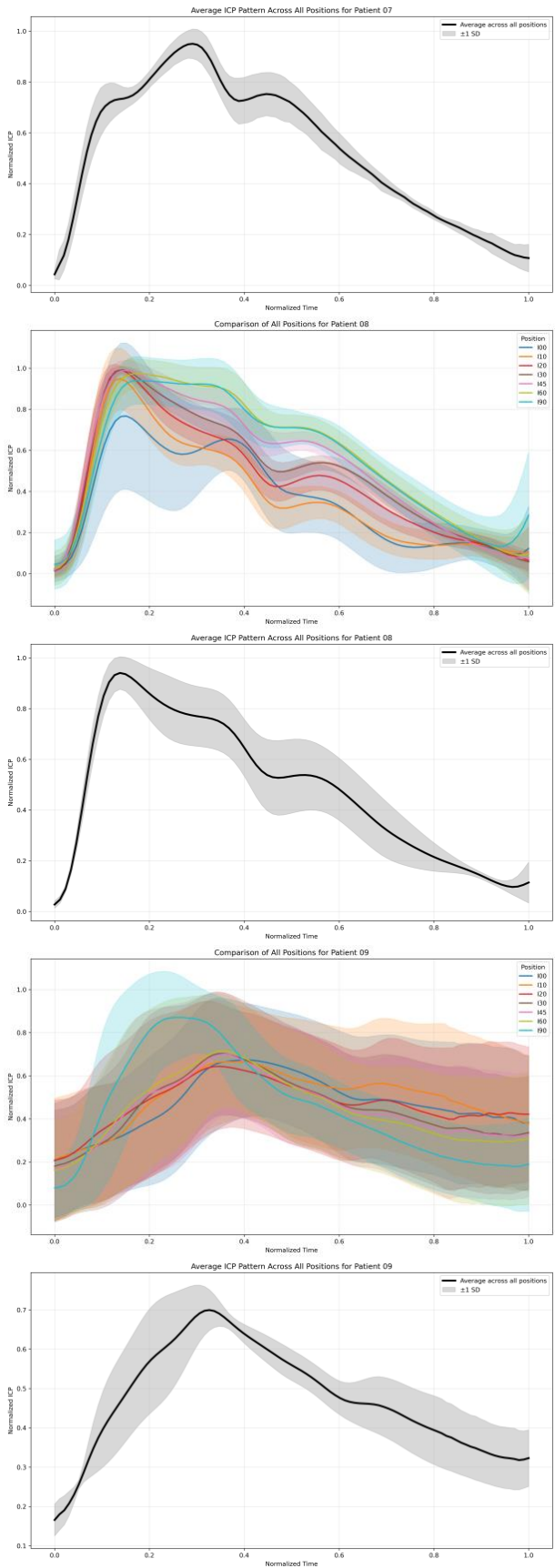

**Fig. 9.** Comparison of all positions for each of 10 patients (top) and average of positions for each patient (bottom). Bold lines are averages, and shaded regions are +/- one standard deviation.

A visual assessment of similarities and differences between individual patients' averaged waveforms is informative, but difficult to scale. To address this, a 1-D convolutional neural network (CNN) was designed to classify individual cardiac cycles into positions. 1-D CNNs have been demonstrated to perform well on time series classification tasks [16], [17]. The network consisted of three 1-D convolutional layers, a 1-D max pooling layer, a dropout layer, and three linear fully connected layers (Fig. 10). The activation function for the forward pass was a leaky rectified linear unit. The CNN was implemented in PyTorch.

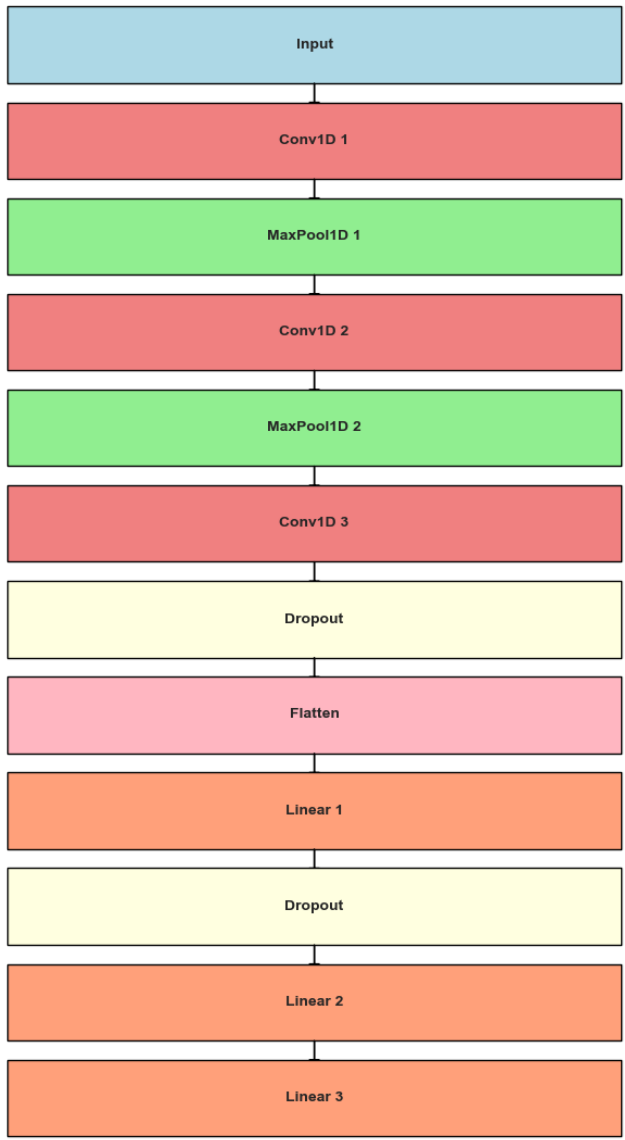

**Fig. 10.** Architectural overview of the neural network.

The data was split into training, validation, and test sets with a ratio of 0.6: 0.2: 0.2. The CNN was trained on a MacBook Air using the Metal Performance Shaders GPU framework. Random seeds were set to 42. Each training loop went for 50 epochs. The CNN used cross entropy loss with a loss rate of 0.001. The NN was trained to differentiate between positions.

The NN was then tuned using an automated hyperparameter optimization approach. Bayesian optimization with Tree-structured Parzen Estimator sampling was conducted using Optuna. Early stopping was implemented with a patience of 50 epochs and a minimum delta of 0.001. 50 trials were conducted with a timeout of 36,000 seconds. Maximum epochs were set to 50 and minimum epochs were set to 10. The maximum number of parameters per model was set to 1,000,000. Evaluation was conducted using cross-validation with stratified splits. The models were evaluated on validation accuracy; the highest validation accuracy was 0.7996. Optimization time was 11161.45 seconds. A model was then trained with the optimized hyperparameters. The optimized hyperparameters and their ranges were: convolutional channels (16-128), kernel size (3-9), fully connected layer size (64-512), dropout rate (0.1-0.5), activation function for convolutional layers (ReLU, LeakyReLU, GELU, Swish), activation function for fully connected layers (ReLU, LeakyReLU, GELU, Tanh), learning rate (1e-4 to 1e-2), optimizers (Adam, AdamW, SGD), batch sizes (16, 32, 64), schedulers (none, cosine annealing, step decay), and weight decay (0 to 1e-3). Optimal parameters were found to be: convolutional channel 1: 64, convolutional channel 2: 64, convolutional channel 3: 64, kernel size for convolutional channel 1: 9, kernel size for convolutional channel 2: 7, kernel size for convolutional channel 3: 3, fully connected layer 1 size: 512, fully connected layer 2 size: 128, dropout rate for convolutional layers: 0.4, dropout rate for fully connected layers: 0.3, activation function for convolutional layers: GELU, activation function for fully connected layers: GELU, batch size: 128, optimizer: AdamW, learning rate: 0.0005, weight decay: 0.001, scheduler: none.

Multiple training runs were conducted to account for the variance and randomness inherent in the non-deterministic nature of neural networks. 35 independent models were trained using the same optimized hyperparameters. For each model, neural network attention was analyzed, i.e., where in the cardiac cycle the neural network was 'paying attention' when learning to classify positions. For each position, a feature importance calculation was performed: perform forward passes through the network; calculate the gradient for each sample with n = 30; take the absolute value of the gradient; create a vector of gradients with length = n; take the mean of the gradients vector to calculate a 'neural importance' score; normalize the neural importance score by the max neural importance score; apply a 1-D Gaussian filter with sigma = 1.0. The calculated importance values over the cardiac cycles were then overlayed on plots of representative individual cardiac cycles (Figure); on plots of the average of all cardiac cycles for each position with +/- 1 SD; and on plots of the average of all cardiac cycles for all positions (Figure). Attention was analyzed for all 35 models.

## III. RESULTS

A single run of the optimized model trained over 50 epochs had a training accuracy of 76.79%, a validation accuracy of 79.24%, and a test accuracy of 80.21% (Fig. 11). The training loss was 0.6125, the validation loss was 0.5633, and the test loss was 0.5431. A confusion matrix for the same run of the optimized model (Fig. 12) shows some overlap between adjacent positions, most likely due to the method used to extract positional data. When changing position, the ICP takes time to adjust to the new position, and judging when the ICP has fully adjusted is a judgement call the extraction personnel must execute. There is no robust process currently in place to clearly differentiate between positions. Best practice is to remove approximately the first minute of data after the patient changed positions, but the amount of time removed varied both within patients and between patients.

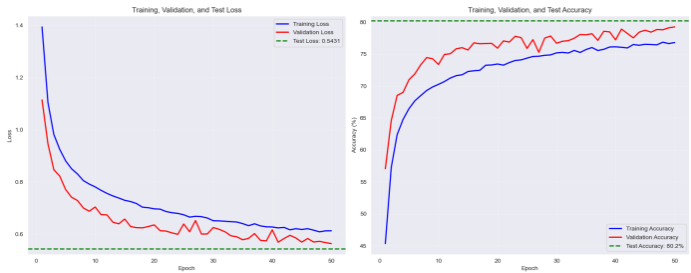

**Fig. 11.** Training, validation and test loss and accuracy for a single run of the optimized model over 50 epochs.

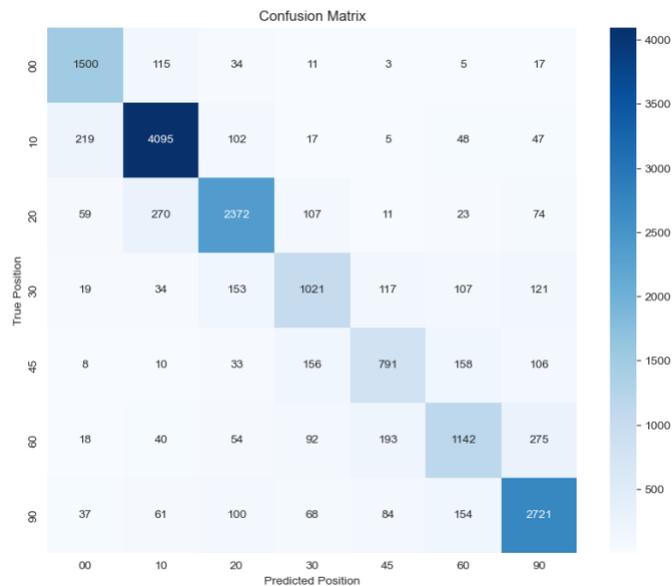

**Fig. 12.** Confusion matrix for a single training run of the optimized model. Overlap between adjacent positions may be a result of the manual data extraction method.

When comparing the averages of all attention patterns between positions in a single run of the model (Fig. 13), the morphology of the attention pattern appears most similar between 10° and 20°. Apparent morphological similarities do not necessarily correspond one-to-one with the results of the confusion matrix, indicating neural attention is not necessarily the only mechanism by which the network classifies cardiac cycles by position. An alternative visualization overlaying the attention maps of each position on each other for a single run (Fig. 14) shows the significant variability in attention between positions.

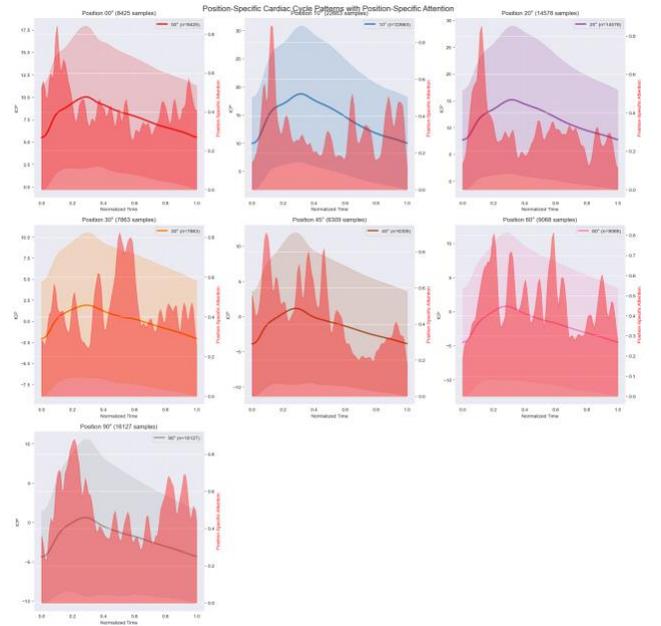

**Fig. 13.** Neural network attention overlaid on average of all cardiac cycles for each position in a single training run of the optimized model. Shaded region indicates +/- one standard deviation.

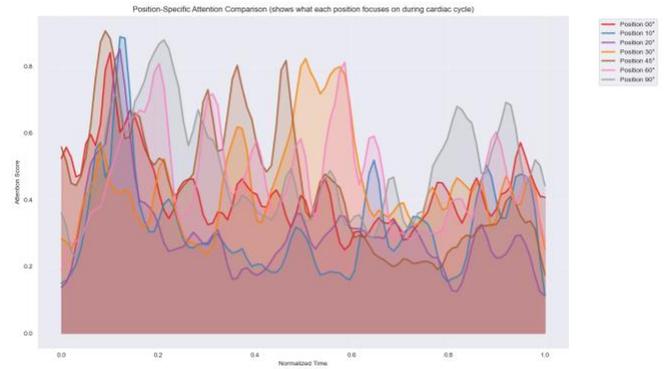

**Fig. 14.** Neural network attention for each position in a single training run of the optimized model.

Visualizing all 35 training runs over each position (Fig. 15) affords a clearer picture of attention mechanisms in ensemble. Visually, 45° appears to have the least variance in the attention map between training runs, and 20° appears to have the most variance. The attention variance map (Fig. 16) seems to support this conclusion. However, quantitative measures of variance are necessary. Overlaying the averages of the attention mechanisms for each position in ensemble (Fig. 17) gives a better visualization of the variance between positions. 45° appears to be the least like all other positions. Across all models, high consensus areas appear between 0 and 0.2 seconds and 0.8 seconds (Fig. 18).

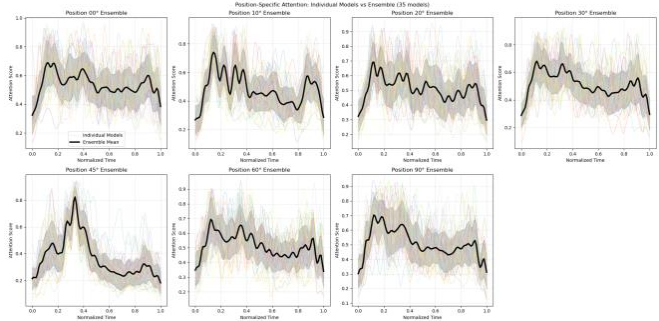

**Fig. 15.** Averaged neural network attention for each position, averaged over 35 training runs, with the shaded region indicating +/- one standard deviation, overlayed on actual importance values from 35 training runs.

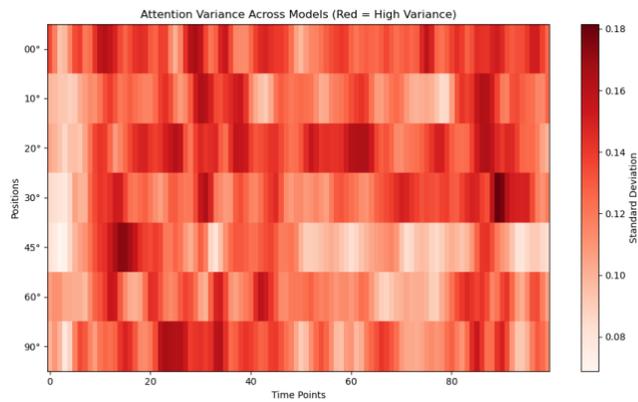

**Fig. 16.** Attention variance over the cardiac cycle for each position, over 35 training runs (bottom).

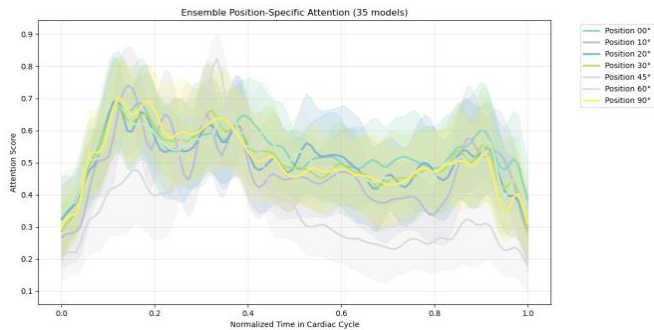

**Fig. 17.** Averaged neural network attention for each position over 35 training runs, with shaded region indicating +/- one standard deviation

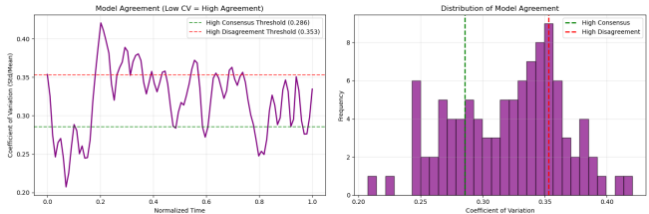

**Fig. 18.** Variance between models shows low consensus.

## IV. DISCUSSION

To interpret the results of neural network attention over different positions, an understanding of the physiological underpinnings of the cardiac cycle and the physiological correlates of positional testing are both necessary. Neural network attention indicates where in the cardiac cycle features are being used to identify position. For example, 45° attention is highest around P2. Other positions focus more on P1, then P2, then at the end of the cardiac cycle. By visual inspection (Fig. 15):

0°: average ICP from $5 - 10$, range from $0 - 17.5$. Highest attention at 0.1, 0.4, 0.9.

10°: average ICP from $10 - 18$, range from $1 - 30$. Highest attention at 0.1, 0.2 − 0.4, 0.9.

20°: average ICP from $8 - 15$, range from -1 − 28. Highest attention at 0.1, 0.3 − 0.4, 0.5, 0.9.

30°: average ICP from $-2 - 2.5$, range from -8.5 − 10.5. Highest attention at 0.1, 0.3, 0.9.

45°: average ICP from -4 − 1, range from -11 − 12. Highest attention at 0.3-0.4.

60°: average ICP from -4.5 − 1, range from -12 − 11. Highest attention at 0.1, 0.3-0.4, 0.9.

90°: average ICP from -4 − 1, range from -10 − 10. Highest attention at 0.1, 0.3, 0.9.

In general, attention seems to focus first on the first inflection point and directly after; second on the second inflection point and directly after (peak); third on the end of the cardiac cycle. For 45°, attention seems to focus more on the second inflection point. Classically, positional changes have been found to be associated with changes in compliance. We may expect to see the most attention focused on P2 as a result. However, this does not appear to be the case, indicating the neural network attention may differ from a physician's attention for the same testing regime.

Positional testing is a single variable that correlates with brain compliance. To obtain a fuller picture of the physiological underpinnings of the cardiac cycle, other variables must be considered using the same backwards approach demonstrated here: identify how features of interest correlate with physiology; assess neural network attention; use the attention to derive information about the physiology indicated by the cardiac cycle. Some examples of other variables of interest are those indicating venous changes, arterial changes, brain compliance, and brain functionality changes such as sleeping vs. awake, sleep stages, standing vs. sitting, heart rhythms (tachycardia and bradycardia), heart rate, cognitive state, hypotensive vs. hypertensive, age, sex, weight; or clinical variables such as diagnosis of normal pressure hydrocephalus, Chiari malformation, idiopathic intracranial hypertension. Once the attention map is generated for each variable, it can be compared between variables. This constitutes an assay approach to discovering the physiological and clinical underpinnings of cardiac cycle morphology. Conversely, classifying the cardiac cycle over these variables can provide clinicians with a diagnostic tool that supplements their current use of ICPm data. Further approaches can be prospective, where specific interventions with known physiological effects are administered, and the same process is then followed. For example, the patient may undergo targeted changes in his or her blood pressure during ICPm. Experimentally validated ICP in animals, where direct experimental physiological and functional changes are possible, and changes can be directly observed and demarcated via cause and effect, may provide better information about physiological underpinnings. More broadly, this approach can be used for different biosignals, and for any signal where the underpinning causes of the signal are not well known. This approach can also be used for multimodal signals to indicate more targeted physiological causes.

A challenge to this approach is the non-deterministic nature of neural networks that can lead to variance in attention. This makes it challenging to draw consistent interpretable features from the feature space, and therefore reproducibility and reliable interpretation of the results can be difficult. In future work, as many training runs as possible should be conducted to ensure the signal-to-noise ratio provides a clear attention map. Recent advances in reducing variance in neural networks may also be beneficial for future improvements to this framework. Additionally, while the current approach allows for a determination of where in the cardiac cycle relevant features may reside, it is not yet understood what the features are that drive this importance metric. Further analysis of cardiac cycles at the location of the increased attention may allow clinicians to better understand the specific indicators within the cardiac cycle delineating different variables such as patient conditions, body position, or physiology. Because the feature layer in a neural network is hidden, and the features cannot be directly extracted and interpreted, a selection and analysis of classical features in regions of interest may be beneficial for future study. Relation extraction using a dual-network-based model may be a promising future approach [19]. This analysis was conducted only on the cardiac cycle. However, as previously discussed, the ICP waveform contains physiologically and clinically relevant information at multiple timescales. An extension to this framework may use analyses at multiple timescales to discover or derive further features. This may require implementation of different neural network architectures like graph neural network [20], [21], [22] to address causal associations [23], transformers [24], state space models [25], [26], or contrastive learning [27]. Comprehensive discussions of models here [28], [29], [30] shows the complexity of time series analysis itself, and of selecting an appropriate model to conduct analyses. A challenge to the analysis of ICP at longer timescales is that as the timescale increases, the conditions under which the ICP was acquired change, requiring careful selection of the data. A patient may be monitored for multiple days, during which they might sleep, sit up, lie down, undergo positional testing, or other activities or experimental interventions. To effectively compare variables between patients and reduce intra-patient variance, selection of overnight time periods when the patient is lying down may be a good starting point to reduce variance.